# Anomalous transport and particle acceleration at shocks

P. Duffy[1], J.G. Kirk[1], Y.A. Gallant[1], R.O. Dendy[2]

[1] Max-Planck-Institut für Kernphysik, Postfach 10 39 80, D-69029 Heidelberg, Germany
[2] UKAEA Government Division, Fusion, UKAEA/Euratom Fusion Association, Culham, Abingdon, Oxon, OX14 3DB, UK

**Abstract.** The theory of first order Fermi acceleration at shocks assumes that particles diffuse due to scattering off slow-moving magnetic irregularities. However, cosmic rays are closely tied to magnetic field lines, and the transport process, particularly across the direction of the field, is likely to be more complicated. To describe cross field transport we employ recent extensions of the Rechester-Rosenbluth theory in localised stochastic regions of magnetic field. During acceleration at a shock and when the motion along the field is diffusive, there is a transition at a critical energy from "sub-diffusive" motion, where the mean square displacement of a particle increases with time as $t^{1/2}$, to compound diffusion, a combined process involving diffusion along a magnetic field which is itself wandering. Requiring this critical energy to be less than the system cut-off in a SNR of radius $R$ places an upper limit on the coherence length, $\lambda$, of the magnetic field for which diffusive shock acceleration can occur at quasi-perpendicular shocks when field line wandering dominates the effective transport of particles: $\lambda < R(U/v)^{1/2}$ where $U$ is the shock speed and $v$ the particle speed. The acceleration rate for particles in the sub-diffusive regime is derived and its implication for the origin of high energy cosmic rays is discussed.

## 1. Introduction

The theory of diffusive particle acceleration at shocks is based on multiple interactions of a particle with a shock front, between which the particle diffuses through the upstream or downstream plasma. In the simplest situations a power-law spectrum of accelerated particles is produced, independent of the diffusion coefficient. This process is thought to be responsible for the acceleration of cosmic rays to energies $\sim 10^{14}$eV in supernova remnants (see Blandford & Eichler 1987 for a review). The maximum energy which can be reached depends on the particle's diffusion coefficient and the age of the shock. Most models of supernova remnants assume gyro-Bohm diffusion (usually referred to as simply "Bohm diffusion"), equivalent to large-angle elastic scattering with a mean-free path equal to the gyro-radius. Many shocks, such as those propagating in the wind of the progenitor star of a supernova explosion, are likely to be quasi-perpendicular, in which case the acceleration rate is determined by diffusion across the magnetic field (Ball & Kirk 1992). If the mean-free path is far greater than the gyro-radius then the spatial diffusion coefficient across the field lines $\kappa_\perp$ can be much smaller than the Bohm value, resulting in a faster acceleration rate (Jokipii 1987). However, it is well-known that turbulence in a magnetised plasma can lead to wandering or braiding of the magnetic field lines: this has been extensively investigated in the context of transport in fusion plasmas but was first studied in connection with the propagation of cosmic rays (Getmantsev 1963, Jokipii & Parker 1969). Clearly this process will influence the transport of charged particles across the average field direction in the vicinity of a shock front (Achterberg & Ball 1994). Here we consider the situation in the light of recent work in transport theory (Isichenko 1991, Rax & White 1992, Chuvilgin & Ptuskin 1993) and find that there exists an energy range in which the transport of energetic particles cannot be described by a diffusion coefficient, but is rather "sub-diffusive" in that the mean square displacement of a particle perpendicular to the magnetic field increases with time as $t^{1/2}$ (cf. Getmantsev 1963). This has a strong influence on the acceleration rate, which we calculate in Sect. 4, using the propagator appropriate to sub-diffusion. At higher energies particle motion across the field becomes diffusive, the "compound diffusion" case, but with a coefficient which combines the effects of scattering off magnetic irregularities with magnetic braiding. In Sect. 5 we discuss the implications of our results for the origin of high energy cosmic rays, and point out the main limitations of our treatment, which applies only in the quasi-linear regime.

## 2. Braided magnetic field

Consider a uniform magnetic field of strength $B$ in the $z$ direction superimposed upon which there is a stationary stochastic magnetic field characterised by a root mean square amplitude $B_1$ and let $b = B_1/B$. If the braiding of the field can be considered quasi-linear, a field line passing through $x = y = z = 0$ wanders such that its distance from the $y$–$z$ plane $x(s)$, as a function of the distance measured along the field line $s$ ($s = 0$ at $z = 0$), follows a Gaussian law: $\langle x^2(s) \rangle = 2 D_M s$ where $\langle \ldots \rangle$ denotes an ensemble average, and $D_M$ is a diffusion coefficient, with the dimensions of a lengthscale. We consider turbulence characterised by the correlation lengths $\lambda_\parallel$ and $\lambda_\perp$ along and perpendicular to the field, and by the Lyapunov length $\lambda_L$ describing the exponential separation of adjacent field lines; i.e. field lines whose initial separation is less than $\lambda_\perp$. Then the diffusion coefficient of the magnetic field is $D_M = b^2 \lambda_\parallel / 4$ and the condition for quasi-linear behaviour is $b \ll \lambda_\perp / \lambda_\parallel$ (Kadomtsev & Pogutse 1979). Neglecting background shear of the magnetic field, quasi-linear theory gives $\lambda_L = \lambda_\perp^2 / (\lambda_\parallel b^2)$ (Isichenko 1991).

## 3. Particle transport

If the particles undergo large-angle elastic scatterings the resulting diffusion coefficients parallel and perpendicular to the field are related to the Bohm value $\kappa_B = \gamma v^2 mc/(3eB)$, for a particle of mass $m$, charge $e$ moving at speed $v$ (and with $\gamma = (1 - v^2/c^2)^{-1/2}$) by

$$\kappa_\parallel = \kappa_B / \epsilon \quad , \quad \kappa_\perp = \epsilon \kappa_B / (1 + \epsilon) \; , \tag{1}$$

where $\epsilon \leq 1$ is the ratio of the collision rate to the gyrofrequency (e.g., Chuvilgin & Ptuskin 1993). In quasi-linear theory, the relationship between these quantities depends on the properties of the turbulence (e.g., Achatz et al. 1991). In particular, $\epsilon$ is in general energy dependent. However, it is usually assumed that both $\kappa_\perp$ and $\kappa_\parallel$ are monotonically increasing functions of energy, and that $\kappa_\perp \ll \kappa_\parallel$. In addition to this diffusion along and across the local magnetic field, a particle's position is also governed by the wandering or braiding of the field. Following Rechester & Rosenbluth (1978), consider a particle moving within a coherent patch of field lines with dimension $\lambda_\perp$ across the field. As the particle moves along the field, the patch is deformed, but maintains constant area, provided the average field strength does not change. The field lines, however, diverge exponentially on the scale of the Lyapunov length $\lambda_L$, so that the patch becomes filamented. After travelling a distance $s$ along the field, the length of a patch filament is roughly $\lambda_\perp \exp(s/\lambda_L)$, so that the typical thickness is $d(s) = \lambda_\perp \exp(-s/\lambda_L)$. Thus, the patch is stretched and the filaments become so thin that the particle can escape into an uncorrelated patch. Escape occurs either when the particle diffuses a distance $d$ across the field, or when $d$ becomes so small that the particle's gyromotion takes it out of the patch (Isichenko 1991). Thus, the escape or decorrelation time $t_d$ is the solution of the equation

$$s(t_d) = \lambda_L \log \left[ \frac{\lambda_\perp}{\mathrm{Max}(r_g, \sqrt{4 \kappa_\perp t_d})} \right] \; . \tag{2}$$

We can now classify the type of transport according to whether the particle is able to travel this distance ballistically, or must diffuse along the field line.

Ballistic motion, for which $s = vt$, requires the particle to propagate for a time given by the solution of Eq. (2) before its parallel velocity is affected by scatterings i.e., $t_d < t_{sc} \equiv \kappa_\parallel / v^2$. According to Eq. (1) this condition implies $r_g > \sqrt{\kappa_\perp t_d}$ and hence $v t_d = \lambda_L \log(\lambda_\perp / r_g)$. In this case diffusive motion in the $x$ direction starts after a time $t = \lambda_\parallel / v$, and continues indefinitely. Thus, the particle suffers not only 'real' cross-field diffusion (for non-zero $\kappa_\perp$) but also transport due to field line wandering:

$$\frac{\langle x^2(t) \rangle}{2t} = \kappa_\perp + D_M v = \kappa_\perp \left( 1 + \frac{t_{sc}}{t_d} \frac{\lambda_\perp^2}{r_g^2} \log(\lambda_\perp / r_g) \right) \tag{3}$$

for $t_d = \lambda_L \log(\lambda_\perp / r_g)/v < t_{sc}$. This formula is valid for all $t > \lambda_\parallel / v$ and yields an effective diffusion coefficient which is large compared to $\kappa_\perp$. The time at which the particle leaves a coherent patch of field plays no role, since diffusion is a Markov process i.e., it can be stopped and restarted in a new patch without any effect on the transport.

If the particle remains within a coherent patch of field for a time $t \gtrsim \kappa_\parallel / v^2$, its motion in $s$ is diffusive: $s(t) = \sqrt{2 t \kappa_\parallel}$. According to Eq. (1) this implies $r_g \lesssim \sqrt{4 \kappa_\perp t_d}$ so that Eq. (2) yields $y = \log(\Lambda / y)$, where $y \equiv \sqrt{2 \kappa_\parallel t_d} / \lambda_L$ and $\Lambda \equiv b^2 \lambda_\parallel / (\epsilon \lambda_\perp \sqrt{2})$. If $\Lambda \lesssim 1$, field-line wandering is unimportant. There is then no anomalous transport regime, and the particle diffuses according to the coefficients of Eq. (1) (Isichenko 1991). Here, however, we are concerned with the opposite case. Assuming $\Lambda \gg 1$, the lowest order solution is $y \approx \log \Lambda$ so that $t_d = \lambda_L^2 (\log \Lambda)^2 / 2 \kappa_\parallel$ which gives two possibilities for motion perpendicular to the mean magnetic field. For times $t < t_d$, the motion has a sub-diffusive component, since the particle is trapped within a coherent patch:

$$\frac{\langle x^2(t) \rangle}{2t} = \kappa_\perp + 2^{1/2} \frac{D_M \kappa_\parallel^{1/2}}{t^{1/2}} \qquad \text{for } t < t_d \tag{4}$$

However, for $t > t_d$, the particle escapes and restarts motion in a new coherent patch. Over such timescales the transport process is Markovian, since memory of the previous patch is lost. The motion is diffusive:

$$\frac{\langle x^2(t) \rangle}{2t} = \kappa_\perp + \frac{2 D_M \kappa_\parallel}{\lambda_L \log \Lambda} = \kappa_\perp \left( 1 + \frac{\Lambda^2}{\log \Lambda} \right) \tag{5}$$

for $t > t_d = (\lambda_L \log \Lambda)^2 / 2 \kappa_\parallel > \kappa_\parallel / v^2$. This is the "collisional" transport regime of Rechester & Rosenbluth (1978) and the process we refer to as "compound diffusion".

## 4. Acceleration at shocks

The existence of different transport regimes has been discussed with reference to cosmic ray propagation, (Chuvilgin & Ptuskin 1993), where the behaviour on short timescales is not important. In the case of particles undergoing acceleration at a shock front, however, these timescales must be compared with the average time spent by a particle in the upstream or downstream medium (the *residence time*, $t_{\rm res}$) before returning to the shock front. For diffusive transport in the coordinate along the normal to the shock front $t_{\rm res} \approx \kappa_{\rm eff}/(Uv)$, where $U$ is a fluid velocity and $\kappa_{\rm eff}$ the effective diffusion coefficient (Toptyghin 1980). Thus, standard diffusion – governed by the coefficients of Eq. (1) – gives a residence time which is an increasing function of particle energy, leading to very slow acceleration at high energy. According to Eq. (5), the effective diffusion coefficient at a perpendicular shock in the compound diffusion regime is, apart from a logarithmic factor, proportional to $\kappa_\parallel$, which increases with energy. We expect this regime to apply at sufficiently high particle energy, since the decorrelation time decreases with increasing energy whereas the residence time increases (see Fig. 1).

Let us assume $\lambda_\perp = \lambda_\parallel = \lambda$, a macroscopic quantity comparable with the scale of the accelerating system – e.g., the radius $R$ of a supernova shock front. Eq. (1) assumes magnetic irregularities of amplitude $\epsilon \approx \langle (\delta B)^2 \rangle / B^2$ on the scale of a gyro radius $r_{\rm g}$, which is much smaller than $\lambda$ (Chuvilgin & Ptuskin 1993). Transport is diffusive for those particles for which $t_{\rm res} > t_{\rm d}$, or, using Eq. (5)

$$\frac{\kappa_\perp}{Uv} \frac{\Lambda^2}{\log \Lambda} > \frac{\lambda_{\rm L}^2 (\log \Lambda)^2}{\kappa_\parallel} \,, \tag{6}$$

for large $\Lambda$, where the diffusion coefficients $\kappa_\parallel$ and $\kappa_\perp$ are monotonically increasing functions of energy. This condition gives a critical energy below which the time a particle is trapped in a coherent field patch $t_{\rm d}$ exceeds the residence time and the transport is sub-diffusive. The highest energy particles accelerated diffusively in a supernova can be found by setting the acceleration time, $t_{\rm acc} \approx \kappa_{\rm eff}/U^2$, equal to the expansion timescale $R/U$. If acceleration to this energy occurs at predominantly perpendicular shocks where field line wandering is important, we arrive at the condition

$$\frac{\kappa_\perp}{U^2} \frac{\Lambda^2}{\log \Lambda} < \frac{R}{U} \,. \tag{7}$$

From Eqs. (6) and (7) one can derive an upper limit

$$\lambda < \left(\frac{U}{v}\right)^{1/2} \frac{R}{(\log \Lambda)^{1/2}} \tag{8}$$

which depends only logarithmically on $b$ and $\epsilon$. If (8) is fulfilled, the highest energy particles can be accelerated diffusively. If it is violated, transport is sub-diffusive at all relevant energies. For energetic particles in supernova remnants $U \ll v$, so this condition places a strong upper limit on the correlation length of the field fluctuations.

A particle residing on either side of a perpendicular shock for a time $t < t_{\rm d}$ will sub-diffuse when the cross-field transport is dominated by field line wandering. Such particles will nevertheless be accelerated as they cross and re-cross the shock front. However, this problem has so far not been addressed, so that it is not a priori obvious that the standard results of diffusive shock acceleration (e.g., the spectrum and spatial dependence of the distribution) are unchanged. Here we restrict ourselves to a discussion of the expected acceleration rate, which is different from that of the standard diffusive picture. In the steady state limit, with $n_{\rm s}(p)$ the number density of particles at the shock with momentum $p$, the flux across the discontinuity is $n_{\rm s}v/4$ where $v$ is the particle speed and the distribution is assumed isotropic. In diffusive shock acceleration there is a lengthscale, $\kappa/U_1$, associated with the upstream distribution which is roughly the distance a particle can diffuse, $x(t) \approx \sqrt{\kappa t}$, before being overtaken by the shock, which is moving at speed $U_1$ relative to the upstream medium. The total number of particles in the upstream medium is then the number density at the shock times this lengthscale, $n_{\rm s}\kappa/U_1$. Dividing by the flux gives $t_{\rm res} = 4\kappa/U_1 v$ as the mean time a particle resides upstream before it enters the downstream region. With a similar expression for particles in the downstream medium, moving at speed $U_2$, which do not escape, the acceleration timescale for diffusive acceleration

$$t_a = \frac{3\kappa}{U_1 - U_2} \left(\frac{1}{U_1} + \frac{1}{U_2}\right) \tag{9}$$

is obtained since a particle's momentum increases by a factor $4(U_1 - U_2)p/3v$ each cycle. In the sub-diffusive case, the distance a particle can travel upstream in time $t$ is now given by $x(t) \approx D_{\rm M}^{1/2} (\kappa_\parallel t)^{1/4}$. The lengthscale of the upstream distribution, as defined above, is then $D_{\rm M}^{2/3} \kappa_\parallel^{1/3}/U_1^{1/3}$. With the assumptions of isotropy and time independence the mean residence time of a particle upstream is roughly $(U_1 D_{\rm M})^{2/3} \kappa_\parallel^{1/3}/U_1 v$ to within a constant factor of order unity. Consequently, as far as the timescales are concerned and to order of magnitude, $(U_1 D_{\rm M})^{2/3} \kappa_\parallel^{1/3}$ plays the role of an effective diffusivity for the non-Markovian transport, i.e. the acceleration timescale is roughly this expression divided by $U^2$. While in the diffusive limit $t_{\rm res} \propto \kappa_\parallel$, the residence time now increases less rapidly with momentum, $t_{\rm res} \propto \kappa_\parallel^{1/3}$ (Fig. 1).

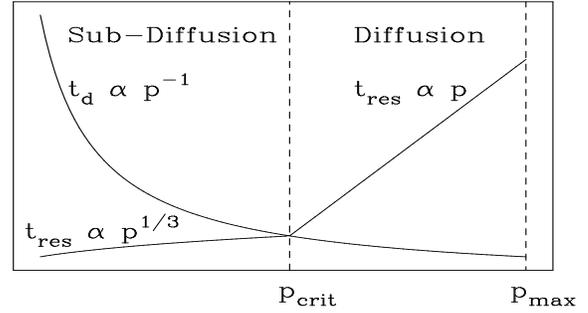

**Fig. 1.** Schematic view of the two regimes for particle acceleration at a quasiperpendicular shock. The decorrelation time $t_{\rm d}$ is a decreasing function of particle momentum $p$ so that particles with momentum below a certain value, $p_{\rm crit}$, sub-diffuse and diffusive shock acceleration is only appropriate for the highest energy particles for which $p_{\rm crit} < p < p_{\rm max}$. The case $\epsilon$ = constant is shown.

This argument can be made more formal by noting that the Gaussian propagator for diffusive processes is replaced by a probability density, $P(x,t)$, of a particle being displaced a distance $x$ in time $t$ which depends on two processes. First there is the probability, $P_{\rm f}(x,s)$, that after a distance $s$ along the field line the perpendicular displacement is $x$. Secondly $P_{\rm p}(s,t)$ is the probability that scattering along the field transports the particle a distance $s$ in time $t$. Then $P(x,t)$ is the product of two Gaussians $P_{\rm f}(x,s)$ and $P_{\rm p}(s,t)$ describing field line wandering and wave-induced diffusion, where $s$ plays a temporal and spatial role respectively, integrated along the magnetic field line;

$$P(x,t) = \int_{-\infty}^{+\infty} ds \; \frac{\exp\left(-x^2/4D_{\rm M}|s|\right)}{\sqrt{4\pi D_{\rm M}|s|}} \frac{\exp\left(-s^2/4\kappa_\parallel t\right)}{\sqrt{4\pi \kappa_\parallel t}} \quad (10)$$

The second moment of this expression gives $\langle x^2(t)\rangle \propto D_{\rm M}\kappa_\parallel^{1/2}t^{1/2}$ as expected. Using the method of steepest descent the integral can be evaluated to give, after normalisation,

$$P(x,t) \approx \alpha \left(D_{\rm M}\kappa_\parallel^{1/2}|x|t^{1/2}\right)^{-1/3} \exp\left(\frac{-\beta|x|^{4/3}}{D_{\rm M}^{2/3}\kappa_\parallel^{1/3}t^{1/3}}\right) \quad (11)$$

(cf. Rax & White 1992) where $\alpha = 2^{-1/3}(3\pi)^{-1/2}$ and $\beta = 3/2^{8/3}$. The mean residence time of a particle upstream of a shock front is found by computing the total time $t_{\rm t}$ spent upstream of the plane $x = U_1 t$ by a particle diffusing according to the propagator (10): $t_{\rm t} = \int_{U_1 t}^\infty dx \int_0^\infty dt\, P(x,t)$. The flux escaping from upstream is $vP(U_1 t, t)/4$ which, when integrated over $t > 0$, gives the average number of times $N_{\rm cr}$ a particle leaves the upstream region. The mean residence time is then

$$t_{\rm res} = t_{\rm t}/N_{\rm cr} = \frac{4(U_1 D_{\rm M})^{2/3}\kappa_\parallel^{1/3}}{U_1 v} \quad (12)$$

in accordance with the physical argument above. The acceleration timescale is then

$$t_{\rm acc} = \frac{3D_{\rm M}^{2/3}\kappa_\parallel^{1/3}}{(U_1 - U_2)}\left(\frac{1}{U_1^{1/3}} + \frac{1}{U_2^{1/3}}\right) \quad (13)$$

with a trivial extension when $D_{\rm M}$ and $\kappa_\parallel$ are discontinuous across the shock. Eq. (13) only holds for particles which do not have time to decorrelate from the field in the vicinity of the shock, i.e. for the lower energy particles in Fig. 1. If we assume $\lambda \approx R$, equating the expansion timescale of a supernova to the sub-diffusive acceleration time indicates that the highest energy to which particles can be accelerated is given by the solution of the implicit equation $\kappa_{\rm B} \approx (RU)/(b^2\Lambda)$ which replaces the usual equation valid for gyro-Bohm diffusion $\kappa_{\rm B} = RU$.

## 5. Discussion

A long-standing problem in the theory of the origin of cosmic rays is that observations demand a single acceleration mechanism for particles between $10\,{\rm GeV}$ and $10^6\,{\rm GeV}$. Diffusive shock acceleration in supernova remnants does not easily produce $10^6\,{\rm GeV}$ particles in the time available. This problem is eased if supernovae accelerate particles at predominantly perpendicular shocks, and if the particle transport at such a shock is governed by the diffusion coefficient $\kappa_\perp$ (Jokipii 1987). Such a picture requires an azimuthal magnetic field configuration around a supernova, which is perhaps not unrealistic (e.g., the spiral field swept out from a rotating star by a stellar wind). We have shown that if particles propagate ballistically over a decorrelation time the effective diffusion coefficient is large compared to $\kappa_\perp$ (Eq. 3). Furthermore, for particles which undergo scattering within a decorrelation time, we have shown that $\kappa_\perp$ is the relevant transport coefficient only if $\Lambda \lesssim 1$, i.e., only if the fluctuations on scales greater than $r_{\rm g}$ are much smaller than those on this scale. Otherwise, field-line wandering controls the particle transport. The diffusive theory of particle acceleration can still be applied to the particles of highest energy, provided that the correlation length of the field fluctuations is sufficiently small. The transport of the highest energy particles is then compound diffusion, governed by an effective diffusion coefficient much larger than $\kappa_\perp$. The problem of diffusively accelerating particles to $10^6$ GeV thus remains.

If the field is coherent on sufficiently long scales, the transport is sub-diffusive and a new theory of particle acceleration is needed. A preliminary calculation for the acceleration rate in this case indicates that rapid acceleration may be possible provided $b^2\Lambda \lesssim 1$. However, the theory we have discussed has several limitations. We have neglected the effects of particle drifts and finite Alfvén velocities on our estimates of the decorrelation time $t_{\rm d}$. These processes may aid a particle in decorrelating from a field fluctuation (Isichenko 1991) and thus make the regime of compound diffusion more accessible. We have also confined our considerations to the quasi-linear regime $b \ll 1$, where it is possible to make simple phenomenological arguments on the lines of Rechester & Rosenbluth (1978) and have assumed an isotropic particle distribution at the shock front. It is unlikely that substantial progress can be made in relaxing these restrictions without undertaking numerical simulations.

**Acknowledgements:** We are grateful for useful discussions with V. Ptuskin. This research was supported in part by the Commission of the European Communities under Contract ER-BCHRXCT940604.